\documentclass[12pt]{article}
\usepackage{graphicx}

\newsavebox{\sboxpubnumber}
\newsavebox{\sboxpubdate}
\newcommand{\pubdate}[1]{\begin{lrbox}{\sboxpubdate}{#1}\end{lrbox}}
\newcommand{\pubnumber}[1]{\begin{lrbox}{\sboxpubnumber}{\begin{tabular}{l} #1 \\
				 \usebox{\sboxpubdate}
				 \end{tabular}}
                           \end{lrbox}
                           \pubblock}
\newcommand{\Title}[1]{\begin{center} {\Large #1 } \end{center}}
\newcommand{\Author}[1]{\begin{center}{ \sc #1} \end{center}}
\newcommand{\Address}[1]{\begin{center}{ \it #1} \end{center}}
\newcommand{\andauth}{\begin{center}{and} \end{center}}
\newcommand{\pubblock}{\rightline{
			\usebox{\sboxpubnumber}}}
\newenvironment{Abstract}{\begin{quotation}  }{\end{quotation}}
\newenvironment{Presented}{\begin{quotation} \begin{center}
             PRESENTED AT\end{center}\bigskip
      \begin{center}\begin{large}}{\end{large}\end{center}
      \end{quotation}}
\newcommand{\Acknowledgements}{\bigskip  \bigskip \begin{center} \begin{large}
             \bf ACKNOWLEDGEMENTS \end{large}\end{center}}
\usepackage{amsmath}

\newcommand{\vc}[1]{\mathbf{#1}}
\newcommand{\mx}[1]{\mathbf{#1}}
\renewcommand{\d}{\mathrm{d}}

\DeclareMathSymbol{\mg}{\mathrel}{symbols}{"1D}  % The >> sign
\newcommand{\Bnabla}{\ensuremath{\boldsymbol{\nabla}}}

\newcommand{\half}{\frac 12 }
\newcommand{\Id}{\mbox{\small 1}\hspace{-3.5pt}\mbox{1}}

\newcommand{\der}{\partial}
\newcommand{\inv}{^{-1}}
\newcommand{\lh}{\left(}
\newcommand{\rh}{\right)}
\newcommand{\labl}[1]{\label{#1}}
\renewenvironment{matrix}{%
  \hskip -\arraycolsep\array{cccccccccc}%
}{%
  \endarray \hskip -\arraycolsep
}
\renewenvironment{pmatrix}{\left(\matrix}{\endmatrix\right)}
\newcommand{\equ}[1]{\begin{gather} #1 \end{gather}}
\newcommand{\equa}[1]{\begin{align} #1 \end{align}}

\newcommand{\tabu}[2]{\begin{tabular}{#1} #2 \end{tabular}}
\newcommand{\pmtrx}[1]{\begin{pmatrix} #1 \end{pmatrix}}
\newcommand{\non}{\nonumber}

\newcommand{\Bsfm}{\mbox{{\sffamily\bfseries m}}}   
\renewcommand{\gg}{\gamma}
\newcommand{\gd}{\delta}

\newcommand{\gf}{\phi}

\newcommand{\gn}{\nu}

\newcommand{\gk}{\kappa}

\newcommand{\gth}{\theta}

\newcommand{\gp}{\pi}

\newcommand{\get}{\eta}

\newcommand{\gG}{\Gamma}

\newcommand{\gO}{\Omega}

\newcommand{\cD}{{\cal D}}
\newcommand{\cH}{{\cal H}}

\newcommand{\cO}{{\cal O}}
\newcommand{\tQ}{{\tilde Q}}

\newcommand{\tge}{{\tilde\epsilon}}

\newcommand{\tgx}{{\tilde\xi}}

\newcommand{\tget}{{\tilde\eta}}

\newcommand{\Bgd}{{\boldsymbol \delta}}

\newcommand{\Bgf}{{\boldsymbol \phi}}

\newcommand{\Bgx}{{\boldsymbol \xi}}

\newcommand{\Bget}{{\boldsymbol \eta}}

\newcommand{\BgO}{{\boldsymbol \Omega}}

\begin{document}

%%%%%%%%%%%%%%%%%%%%%%%%%%%%%%%%%%%%%%%%%%%%%%%%%%%%%%%%%%%%%%%%%%%%%%%%
%%
%% START EDITING HERE!
%%
%%%%%%%%%%%%%%%%%%%%%%%%%%%%%%%%%%%%%%%%%%%%%%%%%%%%%%%%%%%%%%%%%%%%%%%%
\begin{titlepage}
\pubdate{November 2001}                    %fill in the date
\pubnumber{SPIN-2001/28 \\ ITP-UU-01/35 \\ hep-ph/0111370 } %preprint number(s)

\vfill
\Title{Inflationary perturbations with multiple scalar fields}
\vfill
\Author{Bartjan van Tent}
\Address{Spinoza Institute/Institute for Theoretical Physics, 
         Utrecht University \\
         P.O.Box 80.195, 3508 TD Utrecht, The Netherlands}
\vfill
\andauth
\vfill
\Author{Stefan Groot Nibbelink}
\Address{Physikalisches Institut, Universit\"at Bonn \\
         Nu\ss allee 12, D-53115 Bonn, Germany}
\vfill
\begin{Abstract}
The calculation of scalar gravitational and matter perturbations during
multiple-field inflation valid to first order in slow roll is discussed. These
fields may be the coordinates of a non-trivial field manifold and hence have
non-minimal kinetic terms. A basis for these perturbations determined by the
background dynamics is introduced, and the slow-roll functions are generalized
to the multiple-field case. Solutions for a perturbation mode in its three
different behavioural regimes are combined, leading to an analytic expression
for the correlator of the gravitational potential. Multiple-field effects
caused by the coupling to the field perturbation perpendicular to the field
velocity can even contribute at leading order. This is illustrated numerically
with an example of a quadratic potential. (The material here is based on
previous work by the authors presented in hep-ph/0107272.)
\end{Abstract}
\vfill
\begin{Presented}
    COSMO-01 \\
    Rovaniemi, Finland, \\
    August 29 -- September 4, 2001
\end{Presented}
\vfill
\end{titlepage}
\def\thefootnote{\fnsymbol{footnote}}
\setcounter{footnote}{0}

%%%%%%%%%%%%%%%%%%%%%%%%%%%%%%%%%%%%%%%%%%%%%%%%%%%%%%%%%%%%%%%%%%%%%%%%
% The document starts here
%%%%%%%%%%%%%%%%%%%%%%%%%%%%%%%%%%%%%%%%%%%%%%%%%%%%%%%%%%%%%%%%%%%%%%%%
\section{Introduction}

As has been known for a long time, inflation  \cite{Guth,boekLinde} offers a
mechanism for the production of density perturbations,  which are supposed to
be the seeds for the formation of large scale structures in the universe.  This
mechanism is the magnification of microscopic quantum fluctuations in the
scalar fields present during the inflationary epoch into macroscopic matter and
metric perturbations. Also, since a part of the primordial spectrum of density
perturbations is observed in the cosmic microwave background radiation (CMBR),
this mechanism offers one of the most important ways of checking and
constraining possible models of inflation, see e.g.\ \cite{Kinneyetal},
especially when combined with large scale structure data \cite{Hannestadetal}.

There are two important reasons for considering inflation with multiple scalar
fields. The problem of realizing sufficient inflation before a graceful exit 
from the inflationary era and producing the observed density perturbation 
spectrum in a model without very unnatural values of the parameters and initial 
conditions can be solved by the introduction of additional fields.
This is the motivation for hybrid inflation models \cite{Linde,LythRiotto}. The 
other reason is that many 
theories beyond the standard model of particle physics, like grand unification, 
supersymmetry or effective supergravity from string theory, contain a lot of 
scalar fields. Ultimately one would hope to be able to identify those fields 
that can act as inflatons. 
In addition such string-inspired supersymmetric models naturally have 
non-minimal kinetic terms.

The previous two paragraphs outline the motivation for looking at perturbations
in multiple-field inflation. A lot of work in this direction has already been 
done, for example in \cite{PolarskiStarobinsky,GarciaWands,Gordonetal,LythRiotto,
SasakiStewart,NakamuraStewart,PolarskiStar,Langlois,HwangNoh} (more references
can be found in \cite{GNvT}). However, most of the previous literature does not 
consider the most general case, but is usually limited to only two fields and/or 
minimal kinetic terms. The papers \cite{SasakiStewart,NakamuraStewart} do
treat the general case, but the authors do not consider the effect of rotations
of the basis, nor do they work out explicitly the particular contribution to the
gravitational potential. In \cite{GNvT} we provided a general treatment by 
computing the scalar gravitational and matter perturbations to first order in 
slow roll during inflation with multiple real scalar fields that may have 
non-minimal kinetic terms. Which of these fields acts as inflaton during which 
part of the inflationary period is determined automatically in our formalism. 

This paper basically summarizes part of our previous work \cite{GNvT}, 
concentrating on the calculation of the gravitational potential during
inflation, in particular on multiple-field effects like the influence of entropy
perturbations during inflation. Necessary background concepts, like the induced
field basis and the generalized slow-roll functions, are also discussed. The 
result is presented in the form of the gravitational correlator at the time of 
recombination, but for the evolution after inflation only adiabatic
perturbations are considered here. This paper also contains a new result for 
the spectral index $n-1$ of the perturbation spectrum.

\section{Background}

The gravitational background of the universe is described by the flat
Robertson-Walker metric with scale factor $a$:
\equ{
\d s^2 = - \d t^2 + a(t)^2 \d\vc{x}^2 
\qquad \Leftrightarrow \qquad
\d s^2 = a(\eta)^2 \lh -\d\eta^2 + \d\vc{x}^2 \rh.
}
The comoving time $t$ and conformal
time $\eta$ are related by $\d t = a \, \d\eta$. Derivatives
with respect to $t$ and $\eta$ are denoted by a dot and a prime, respectively;
the associated Hubble parameters are $H\equiv\dot{a}/a$ and $\cH\equiv a'/a=aH$.
Another useful variable is the number of e-folds $N$, defined by
\(
a(N) = a_0 \exp(N),
\)
which can be considered as a time variable as well:
\(
\d N = H \d t = \cH \d\eta.
\)

For the matter content of the universe we assume an arbitrary number of real 
scalar fields, which are the components of a vector $\Bgf$. Apart from a
generic potential $V(\Bgf)$ we allow for the possibility of non-minimal kinetic
terms, encoded by a field metric $\mx{G}$. In other words, the scalar fields
may be the coordinates of a real field manifold with a non-trivial metric 
$\mx{G}$.
This is a common situation in for example supergravity models, where the
(complex) scalar  fields parameterize a so-called K\"ahler manifold with a
metric that is the  second mixed derivative of the K\"ahler potential. Since a
complex field can always be written in terms of two real fields, our treatment
with real scalars is sufficiently general to be easily applicable to these
special manifolds.

The Einstein and field equations lead to the following equations of motion for
the homogeneous background:
\equ{
H^2 = \frac{\gk^2}{3} \lh \half |\dot{\Bgf}|^2 + V \rh,
\qquad
\dot{H} = -\half \gk^2 |\dot{\Bgf}|^2,
\qquad
\cD_t \dot{\Bgf} + 3 H \dot{\Bgf} + \mx{G}\inv \Bnabla^T V = 0.
}
Here we have defined the covariant time derivative on a vector in field space as
\(
\cD_t A^a = \dot{A}^a + \gG^a_{bc} \dot{\gf}^b A^c.
\)
(Indices $a,b,\ldots$ are used for components in field space and $\gG^a_{bc}$ is
the connection associated with the field metric $\mx{G}$.)
The $\Bnabla$ is used for covariant derivatives with respect to the fields:
$(\Bnabla V)_a = \der V / \der \gf^a$. The length of a vector is given by
$|\vc{A}|=\sqrt{\vc{A}\cdot\vc{A}}$, with the inner product defined by
$\vc{A}\cdot\vc{B}=\vc{A}^\dag \vc{B}=\vc{A}^T \mx{G} \vc{B}$.
The quantity $\gk$ is the inverse reduced Planck mass: 
$\gk^2 \equiv 8\pi G = 8\pi/M_P^2$.

The background field dynamics induce a prefered basis on the field manifold.
The first unit vector $\vc{e}_1$ is given by the
direction of the field velocity $\dot{\Bgf}$. The second unit vector 
$\vc{e}_2$ points in the direction of that part of the field acceleration 
$\cD_t \dot{\Bgf}$ that is perpendicular to the first unit vector $\vc{e}_1$.
Hence:
\equ{
\vc{e}_1 = \frac{\dot{\Bgf}}{|\dot{\Bgf}|},
\qquad \vc{e}_2 = \frac{\mx{P}^\perp \cD_t \dot{\Bgf}}
{|\mx{P}^\perp \cD_t \dot{\Bgf}|},
\qquad\qquad \mx{P}^\perp = \mx{1} - \mx{P}^\parallel = \mx{1} 
- \vc{e}_1^{\;} \vc{e}_1^\dag,
}
with $\mx{P}^\parallel$ the projection on the direction $\Dot{\Bgf}$.
This Gram-Schmidt orthogonalization procedure can be continued to construct the
remaining basis vectors \cite{GNvT}, but the first two are the only ones we 
need here.

We can now define the following multiple-field slow-roll functions:
\equ{
\tge = - \frac{\dot{H}}{H^2}, 
\qquad\qquad \tilde\Bget = \frac{\cD_t \dot{\Bgf}}{H |\dot{\Bgf}|},
\qquad\qquad \tilde\Bgx = \frac{\cD_t^2 \dot{\Bgf}}{H^2 |\dot{\Bgf}|},
\labl{slowrollfun}
}
of which we can take components with respect to the basis defined above, for
example $\tget^\parallel = \vc{e}_1 \cdot \tilde\Bget$, 
$\tget^\perp = \vc{e}_2 \cdot \tilde\Bget$ and
$\tgx^\parallel = \vc{e}_1 \cdot \tilde{\Bgx}$. 
Again, one can easily define slow-roll functions of higher order \cite{GNvT}.
With these definitions as short-hand notation we can rewrite the background
equations of motion in the following form, which is still exact:
\equa{
H & = \frac{\gk}{\sqrt{3}} \sqrt{V} \lh 1 -\frac{1}{3} \tge \rh^{-1/2},
\labl{Friedmanniter}\\
\Dot{\Bgf} + \frac{1}{\gk \sqrt{3}} \frac{1}{\sqrt{V}} \, \mx{G}\inv \Bnabla^T V
& = -\sqrt{\frac{2}{3}} \sqrt{V} \frac{\sqrt{\tge}}{1-\frac{1}{3}\tge}
\lh \frac{1}{3} \tilde\Bget + \frac{\frac{1}{3} \tge \, \vc{e}_1}
{1+\sqrt{1-\frac{1}{3}\tge}} \rh.
\labl{eqmotbackiter}
}
The assumption that the slow-roll functions $\tge$ and $\tilde{\Bget}$ are 
(much) smaller than unity is called the slow-roll approximation.
(The second order slow-roll function $\tilde{\Bgx}$ is assumed to be of an 
order comparable to $\tge^2$, $\tge\tget^\parallel$, etc.) If this assumption
is valid, we can use expansions in powers of these slow-roll functions to 
estimate the relevance of various terms in a given expression. For example, to 
first order the Friedmann equation \eqref{Friedmanniter} is approximated by
replacing $(1-\tge/3)^{-1/2}$ by $(1+\tge/6)$.
The background field equation up to and including first order is given by 
\eqref{eqmotbackiter} with the right-hand side set to zero, as all 
those terms are order $3/2$ or higher.
%An important point to note is that even taking this slow-roll approximated field
%equation and assuming a trivial field metric, the component field equations will
%still in general be coupled by means of the factor $\sqrt{V}$. Hence the 
%multiple-field case is intrinsically more difficult than the single-field one.

\section{Perturbations}

On top of the homogeneous background treated in the previous section there are
small quantum fluctuations. We consider only scalar perturbations and write the 
matter and metric perturbations as follows \cite{Mukhanovetal}
\equ{
\Bgf^{\mathrm{full}}(\eta,\vc{x}) = \Bgf(\eta) + \Bgd\Bgf(\eta,\vc{x}),
\non\\
g_{\mu\nu}^{\mathrm{full}}(\eta,\vc{x}) 
= g_{\mu\nu}(\eta) + \gd g_{\mu\nu}(\eta,\vc{x}) =
a^2(\eta) \pmtrx{-1 & 0\\ 0 & \gd_{ij}}
- 2 a^2(\eta) \Phi(\eta,\vc{x}) \pmtrx{1 & 0\\ 0 & \gd_{ij}}.
}
All equations are linearized with respect to the perturbation quantities.
The gravitational potential $\Phi(\eta,\vc{x})$
describes the scalar metric perturbations and is a quantity we are interested
in, as it is related to the temperature fluctuations in the cosmic microwave
background radiation by means of the Sachs-Wolfe effect.

Instead of using $\Bgd\Bgf$ and $\Phi$ to describe the perturbations, it turns
out to be more convenient to use the so-called generalized Mukhanov-Sasaki
variables $\vc{q}$ and $u$, as well as a short-hand notation $\gth$, defined by
\equ{
u = \frac{\Phi}{\gk \sqrt{2} \, H \sqrt{\tge}},
\qquad\qquad
\vc{q} = a \lh \Bgd\Bgf + \frac{\Phi}{\cH} \, \Bgf' \rh,
\qquad\qquad
\gth = \frac{\gk}{\sqrt{2}} \frac{1}{a \sqrt{\tge}}.
}
These redefinitions remove first order time derivatives, so that it is easier to
understand the behaviour of the solutions, and are necessary for quantization.
The perturbation equations in terms of spatial Fourier modes $\vc{k}$ now read
\cite{GNvT}
\equ{
u_{\vc k}'' + \lh k^2  - \frac{\gth''}{\gth}\rh u_{\vc k}
=  \cH \tget^\perp \vc e_2 \cdot \vc q_{\vc k},
\qquad\qquad
\cD_\eta^2 \vc{q}_{\vc k} + ( k^2 + \cH^2 {\BgO}) 
\vc{q}_{\vc k} = 0,
\non\\
{\BgO} \equiv \frac{1}{H^2} \tilde{\mx M}^2
- (2 - \tge) \Id 
-  2 \tge \Bigl( ( 3 + \tge) \mx P^\parallel 
+ \vc e_1^{\;} \tilde{\Bget}^\dag + \tilde{\Bget} \vc e_1^\dag \Bigr).
\labl{equ}
}
Here the effective mass matrix $\tilde{\mx{M}}^2$ is defined by
\(
\tilde{\mx M}^2 \equiv \mx G\inv \Bnabla^T \Bnabla V 
- \mx R (\Dot{\Bgf},\Dot{\Bgf}),
\)
with $\mx{R}$ the curvature tensor on the field manifold,
$[\mx{R}(\vc{B},\vc{C})\vc{D}]^a = R^a_{\; bcd} B^b C^c D^d$.
From its definition we can derive that $\gth''/\gth$ is exactly given by
\(
\gth''/\gth = \cH^2 ( 2\tge + \tget^\parallel 
+ 2 (\tget^\parallel)^2 - (\tget^\perp)^2 - \tgx^\parallel ).
\)
We see that in the multiple-field case the redefined gravitational potential $u$
is coupled to the field perturbation in the $\vc{e}_2$ direction. 
In the literature (see e.g.\ \cite{Gordonetal}) perturbations in the $\vc{e}_1$
direction are called adiabatic perturbations, while perturbations in the other
directions are called entropy perturbations. This coupling of the gravitational
potential to the entropy perturbations
is suppressed by the slow-roll function $\tget^\perp$, so one could a priori
expect the contribution of this inhomogeneous term to be only important at first
order. However, because of integration interval effects it turns out that it can
contribute even at leading order, as shown in the example in
section~\ref{example}.

To quantize the perturbations, we explicitly choose to work in the basis 
$\{\vc{e}_1,\vc{e}_2,\ldots\}$ defined in the previous section, denoting 
vectors in that basis with non-bold symbols: 
$q^T = (q_n) = (\vc{e}_n\cdot\vc{q})$.
Although this has several advantages, most importantly it results in a standard
canonical normalization of $\half (q')^T q'$ in the Lagrangean, independent of 
the field metric $\mx{G}$, making quantization straightforward. Of course there 
is the price that terms with $Z$ (see below) appear, but these can be dealt with.
We can write
\(
\hat{q}(\eta) = Q(\eta) \hat{a}^\dag + Q^*(\eta) \hat{a},
\)
with constant creation and annihilation operator vectors $\hat{a}^\dag$ and
$\hat{a}$ and a matrix function $Q(\eta)$ that satisfies the classical equation
of motion.
Finally we perform a rotation $Q=R\tQ$ to simplify this equation of motion and 
find:
\equ{
\tQ''  + ( k^2  + \cH^2  \tilde{\gO} ) \tQ = 0,
\qquad
R' + \cH Z R = 0,
\qquad
Z_{mn} = \frac{1}{\cH} \, \vc{e}_m \cdot \cD_\eta \vc{e}_n,
\labl{eqtq}
}
with $\tilde{\gO} = R\inv \gO R$.
The matrix $Z$ is anti-symmetric, first order in slow roll and only non-zero for
$m=n \pm 1$. The presence of this matrix is caused by the fact that the basis
vectors are in general not static in field space because of the background 
dynamics. Working out the definitions, one can easily show that 
$Z_{21}=\tget^\perp$; general expressions are given in \cite{GNvT}.

Considering the equations \eqref{equ} and \eqref{eqtq} for $u$ and $\tQ$ and
realizing that $\cH$ grows rapidly during inflation, while $k$ is a constant per
mode, we see that their solutions change dramatically around the time 
$\get_\cH$ when a scale crosses the Hubble scale (`passes through the horizon'),
defined by $\cH(\get_\cH) = k$. Hence there are three regions 
of interest: sub-horizon ($\cH \ll k$), transition ($\cH \sim k$) and
super-horizon ($\cH \mg k$). 
(Notice that we are considering a single, though arbitrary, mode $\vc{k}$ here,
on which the resulting expressions will depend.)
The equations in the sub-horizon region are easily
solved, but the oscillatory solutions there are irrelevant for the correlators 
we are interested in.

In the transition region we can only solve the equations analytically if we make
the additional assumption that the slow-roll functions are constant. Since the
derivatives of the slow-roll functions are one order higher in slow roll, this
is a consistent approximation to first order, provided that the transition
region is small enough. The solution for $Q$ valid to first order in slow roll 
in the neighbourhood of $\eta_\cH$, taking into account the correct initial 
conditions, can be written in terms of a Hankel function:
\equ{
Q(\eta) = \sqrt {\frac {\gp\eta}{4} }\,  H_{\gn_\cH}^{(1)}(k\eta),
\qquad
\gn_\cH = \frac{3}{2}\Id + \gd_\cH,
\qquad
\gd_\cH = \tge_\cH \Id - \frac{\tilde{M}_\cH^2}{3 H_\cH^2} 
+ 2 \tge_\cH \, e_1^{\;} e_1^T.
\labl{Qtrans}
}
Here the additional assumption has been made that also those components of the
matrix $\tilde{M}_\cH^2/H_\cH^2$ that cannot be expressed in terms of the
slow-roll functions defined in \eqref{slowrollfun}, are nonetheless of first
order in slow roll. In particular this puts constraints on the curvature tensor
of the field manifold.
The matrix $Z$ appearing in equation \eqref{eqtq} for $R$ is a possible source
of multiple-field effects. However, it turns out that in generic
situations the effects of the rotation of the basis during the transition 
region encoded by $Z$  are beyond
first order so that it does not appear in the first order result \eqref{Qtrans},
although it might be important if $\tget^\perp$ peaks around $\eta_\cH$.

Solutions for $u$ and $Q$ can also be determined analytically to first order in
the super-horizon region. However, matching the solutions in the super-horizon
region to those in the transition region to determine the constants of
integration is not trivial, as there is no region of overlap where both
solutions are valid to first order. Hence the standard method of continuously
differentiable matching at a certain time scale cannot be applied. In particular
one should not simply match at $\eta_\cH$, as the assumption of neglecting $k$, 
that is necessary to obtain the analytic super-horizon solution, is not valid 
there.

We solved this problem in \cite{GNvT} by setting up series expansions in 
$k\eta$ for both the 
super-horizon and transition solutions. We found that the leading order powers 
in these asymptotic expansions can exactly be identified to first order, for 
both the decaying and non-decaying independent solutions, which can already at
zeroth order be distinguished from each other. Since the transition and
super-horizon solutions are approximations of the complete solution
of the same equation of motion, the coefficients in front of the series have to 
be equal as well. In this way the solution in the super-horizon region is 
determined completely to first order. 

The final result for $u$ at later times in the super-horizon region
(i.e.\ neglecting terms that are suppressed by the scale factor) to first order
in slow roll is \cite{GNvT}: 
\equ{ 
u_{\vc k}(\eta) = u_{P\, \vc k}^{\;} +
D_{\vc{k}} \gth \int_{\eta_\cH}^\eta \frac{\d\eta'}{\gth^2}, 
\qquad  
u_{P\, \vc{k}} = \gth \int^\eta_{\eta_\cH} \frac{\d\get'}{\gth^2}  
\int^{\eta'}_{\eta_\cH} \d \get'' \, \cH \gth \tget^\perp q_{2\, \vc{k}},  
}
with  
\( 
D_{\vc{k}} = - \half \frac{e^{-i\pi\gd_\cH}}{i\sqrt{2k}} \, \gth_\cH 
e_1^T \lh (1- \tge_\cH) \Id  + B \gd_\cH \rh \hat{a}^\dag + \mbox{c.c.} 
\) 
and $u_P$ the particular solution caused by the inhomogeneous source term in the
equation of motion \eqref{equ} for $u$ and 
$B \equiv 2-\gg_E -\ln 2 \approx 0.7297$.
Switching back to the real gravitational potential $\Phi$ we have the complete
solution at the end of inflation, so including the effect of the perpendicular
field or entropy perturbations {\em during} inflation. Considering here for 
simplicity only adiabatic perturbations {\em after} inflation we can compute 
the vacuum correlator of the gravitational potential at the time of 
recombination when the CMBR was formed \cite{GNvT}: 
\equa{
|\gd_{\vc{k}}|^2 \equiv \frac{k^3}{2\pi^2} \langle \Phi_{\vc{k}}^2 
\rangle_{t_{rec}} 
= \frac{9}{25} \, \frac{\gk^2}{8 \pi^2}
\frac{H_\cH^2}{\tge_\cH} \Bigl [ & (1-2\tge_\cH)(1 + U_{P\, e}^T U_{P\, e})
\non\\
& + 2 B \lh (2\tge_\cH + \tget^\parallel_\cH)
+ 2 \tget^\perp_\cH e_2^T U_{P\, e} + U_{P\, e}^T \gd_\cH U_{P\, e} \rh \Bigr ].
\labl{CorrelatorPhi}
}
Here $U_{P\, e} = U_P(\eta_e)$, with $\eta_e$ the end time of inflation, is
defined by
\equ{
U_P^T = 2 \sqrt{\tge_\cH} \int_{\eta_\cH}^{\eta} \d\eta' \, \cH
\frac{\tget^\perp}{\sqrt{\tge}} \, \frac{a_\cH}{a} \, e_2^T Q Q_\cH\inv
\labl{defUPe}
}
and we have used the fact that $U_P$ has no component in the $e_1$ direction.
More information on $U_P$, in particular on how to rewrite it in terms of
background quantities only, can be found in \cite{GNvT}.
Apart from this amplitude we can also compute the slope of the spectrum. This
spectral index can be determined one order further, but most important is its 
leading (first) order part:
\equ{
n-1 \equiv \frac{\der \ln |\gd_k|^2}{\der \ln k}
= - 4 \tge_\cH - 2 \tget^\parallel_\cH
+ \frac{2 U_{P\, e}^T (2 \tge_\cH + \tget^\parallel_\cH - \gd_\cH) U_{P\, e}
- 4 \tget^\perp_\cH e_2^T U_{P\, e}}{1 + U_{P\, e}^T U_{P\, e}}.
\labl{slope}
}

The explicit multiple field terms in the results for $|\gd_{\vc{k}}|^2$ and
$n-1$ are the contributions of the terms with
$U_{P\, e}$ and $\tget^\perp_\cH$, which are absent in the single-field case. 
Since they are both to a large extent determined by $\tget^\perp$, we see that 
the behaviour of $\tget^\perp$ during the last 60 e-folds of inflation is 
crucial to determine whether multiple-field effects are important. 

\section{Example}
\labl{example}

In this section the example of a quadratic potential on a flat field manifold 
is briefly discussed to illustrate the theory. In this case it is possible
to compute $U_{P\, e}$ explicitly and we show that it can indeed contribute
already at zeroth order to the gravitational correlator, despite the slow-roll
factor $\tget^\perp$.

The potential is $V = \half \gk^{-2} \, \Bgf^T \Bsfm^2 \Bgf$ with $\Bsfm^2$ 
a general symmetric mass matrix given in units of the Planck mass $\gk^{-1}$.
With a trivial field metric and initial condition $\Bgf(0)=\Bgf_0$ we can write
the general first order slow-roll solution for $\Bgf$ in terms of a single
scalar function $\psi(t)$. Even without knowing this function explicitly we can
determine $U_{P\, e}$ \cite{GNvT}:
\equ{
\Bgf(t) = e^{-\half \mx{m}^2 \psi(t)} \Bgf_0,
\qquad\qquad
U_{P\, e}^T = - \frac{\gk \sqrt{\tge_\cH}}{\sqrt{2}} \, \gf_\cH^T P_1^\perp.
\labl{UPevb}
}
This expression for $U_{P\, e}$ assumes slow roll, but is valid at the end of
inflation provided that $\tget^\perp$ goes to zero.

For the case of two fields with masses $m_1 = 1 \cdot 10^{-5}$ and 
$m_2 = 2.5 \cdot 10^{-5}$ and initial conditions $\gf_1 = 20$ and $\gf_2 = 25$
in Planck units, the results for the slow-roll functions and $U_P$ as a function
of the number of e-folds $N$ are plotted in figure~\ref{fig1}.
We see that in this case $\tget^\perp$ indeed goes to zero at the end of
inflation after reaching a maximum during the last 60 e-folds, and $U_P$ 
reaches a constant value long before a possible break-down of slow roll.
Values for the gravitational correlator amplitude $|\gd_{\vc{k}}|^2$ and
spectral index $n-1$ are given in table~\ref{tab1}.

%%%%%%%%%%%%%%%%%%%%%%%%%%%%%%%%%%%%%%%%%%%%%%%%%%%%%%%%%%%%%%%%%%%%%%%%
%%
%%   use this format to include an .eps figure into your paper
%%
\begin{figure}[tb]
    \centering
    \includegraphics[width=7.5cm]{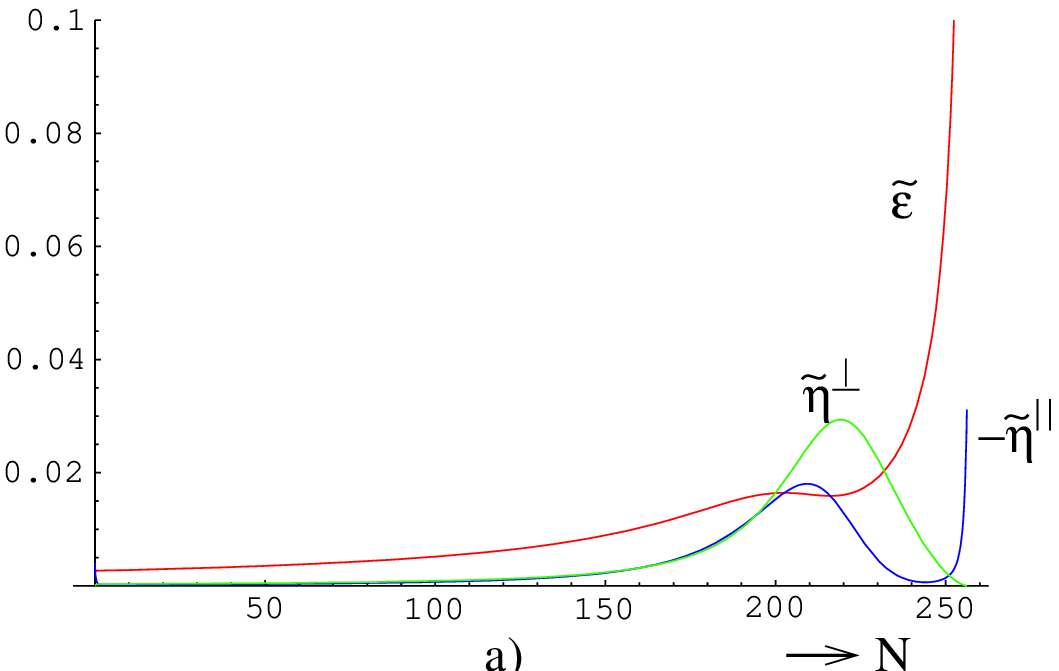}
    \hspace{0.5cm}
    \includegraphics[width=7.0cm]{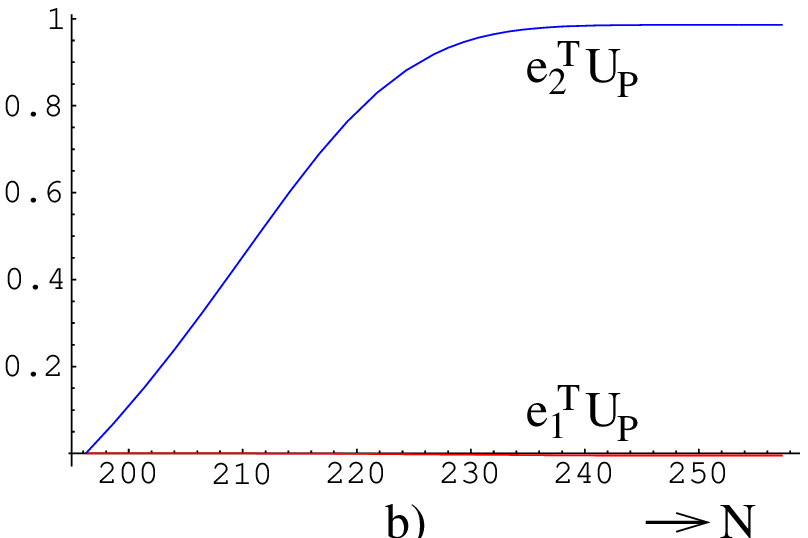}
    \caption[fig1]{\sf a) The slow-roll functions and b) the particular 
    contribution $U_{P}$ to the gravitational correlator
as a function of the number of e-folds $N$ in the model with two fields on 
a flat manifold with a quadratic potential with masses $m_1 = 1 \cdot 10^{-5}$, 
$m_2 = 2.5 \cdot 10^{-5}$ and initial conditions $\gf_1 = 20$, $\gf_2 = 25$.}
    \labl{fig1}
\end{figure}
%%%%%%%%%%%%%%%%%%%%%%%%%%%%%%%%%%%%%%%%%%%%%%%%%%%%%%%%%%%%%%%%%%%%%%%%

\begin{table}[tb]
\small
\centering
\tabu{l||c|c|c||c|c}{
& $|\gd_\vc{k}|^2$ & Contribution & Error
& $n-1$ & Contribution \\ 
\hline
Homogeneous & $1.55 \cdot 10^{-9}$ & 0.505 & 0.0001 & $-0.0377$ & 0.584\\
Particular & $1.52 \cdot 10^{-9}$ & 0.495 & 0.0006 & $-0.0269$ & 0.416\\ 
\hline
Total & $3.08 \cdot 10^{-9}$ & 1 & 0.0003 & $-0.0646$ & 1\\
}
\caption[tab1]{\sf The amplitude $|\gd_{\vc{k}}|^2$ and slope $n-1$ of the 
vacuum correlator of the gravitational potential
are separated into a purely homogeneous and a (mixed) particular part.
Given are their values and their relative contributions to the total
according to our analytical slow-roll results \eqref{CorrelatorPhi} and
\eqref{slope} combined with \eqref{UPevb}. For the amplitude we also calculated
exact numerical results and give here the relative error between those and our 
analyical approximations.} 
\labl{tab1}
\end{table}

The results for the amplitude and the slope are split into a homogeneous part 
(all terms without $U_{P\, e}$) and a particular part (the rest, including 
mixing terms). 
Everything is evaluated for the mode $k$ that crosses the horizon 60 e-folds
before the end of inflation.
From the column giving the relative errors between our first order analytical
results \eqref{CorrelatorPhi} and \eqref{UPevb} on the one hand,
and the exact numerical result on the other, we see that these results
agree with our claim that we computed the correlator to first order 
in slow roll: the relative errors are (much) smaller than $\cO(\tge_\cH)$. 
We also see that our slow-roll approximation for $U_{P}$ is indeed still very 
good at the end of inflation, as indicated by the small error in the particular
part.
From the other columns we see that the particular solution terms are 
responsible for almost half the total result in this model, both for the
amplitude and the spectral index. Hence neglecting 
these terms to leading order, which might naively be done because they couple 
with a $\tget^\perp$ in \eqref{equ}, can be dangerous.

\section{Conclusions}

In this paper we have given a summary of our
general treatment \cite{GNvT} for scalar perturbations on a flat 
Robertson-Walker spacetime in the presence of an arbitrary number of scalar 
fields that take values on a curved field manifold during slow-roll inflation. 
These are the kind of systems that one typically obtains from (string-inspired)
high-energy models. Here we concentrated on the calculation of the vacuum
correlator of the gravitational potential to first order in slow roll, which is
related to the temperature fluctuations that are observed in the CMBR.

A discussion of the background served  as the foundation for this
analysis. The background field dynamics naturally induce an orthonormal basis 
$\{\vc e_1, \vc e_2,\ldots\}$ on the tangent bundle of the field manifold. 
This makes a separation 
between effectively single-field and truly multiple-field contributions 
possible and is a necessary ingredient for correct quantization of the
perturbations.
We also modified the definitions of the well-known  slow-roll parameters to
define slow-roll functions in terms of derivatives of  the Hubble parameter and
the background field velocity for the case of multiple scalar field inflation.
These slow-roll functions are vectors, which can be decomposed in  the basis
induced by the field dynamics. For example, the slow-roll function 
$\tget^\perp$ measures the size of the field acceleration perpendicular to the 
field velocity. Because we did not make the assumption that slow roll is valid 
in the definition of the slow-roll functions,  it is often possible to identify 
these slow-roll functions in exact equations of motion and make decisions about
neglecting some of the terms. 

We generalized the combined
system of gravitational and matter perturbations of Mukhanov et al.\
\cite{Mukhanovetal} by defining the Mukhanov-Sasaki variables 
as a vector on the scalar field manifold. The gravitational potential only 
couples to the scalar field perturbation in the direction $\vc e_2$ with a 
slow-roll factor $\tget^\perp$. First order solutions were found in the three 
different regimes that reflect the change of behaviour for a given mode when
it crosses the Hubble scale. Using a procedure of analytically identifying
leading order terms in asymptotic expansions in $k\eta$ we were able to relate
the solutions in the different regions and find the complete first order result
for the gravitational potential at the end of inflation. 
Considering only adiabatic
perturbations after inflation we could give this result in terms of the vacuum
correlator of the gravitational potential at the time of recombination.
We also determined the spectral index $n-1$.

Multiple-field effects are important in the adiabatic perturbations at the time 
of recombination if $\tget^\perp$ is sizable during the last 
60 e-folds of inflation. The most important source of multiple-field effects is the 
particular solution of the gravitational potential caused by the perpendicular
field perturbations during inflation. We found in our numerical 
example of multiple scalar fields on a flat manifold with a quadratic potential
that this term can contribute already at leading order, even though it enters 
with a slow-roll factor in the equation of motion.
This is true for both the amplitude and the slope of the spectrum.

\Acknowledgements

This work is supported by the European Commission RTN
programme HPRN-CT-2000-00131.
S.G.N.\ is also supported by  
piority grant 1096 of the Deutsche Forschungsgemeinschaft
and European Commission RTN programme HPRN-CT-2000-00148/00152.

\end{document}